\documentclass{article}
\usepackage{arxiv}
\usepackage{hyperref}
\usepackage{latexsym}
\usepackage{url}
\usepackage[utf8]{inputenc}
\usepackage{booktabs}
\usepackage{amsmath}
\usepackage{amsfonts}
\usepackage{amssymb}
\usepackage[inline]{enumitem}
\usepackage{enumitem}
\usepackage{mathbbol}
\usepackage[T1]{fontenc}
\usepackage{mathtools}
\usepackage{multirow}
\usepackage{natbib}
\usepackage{rotating}
\usepackage{subcaption}
\usepackage{graphicx}
\usepackage{etaremune}
\usepackage{authblk}
\usepackage{tikz}

\newcommand{\ie}{\emph{i.e.}}
\newcommand{\eg}{\emph{e.g.}}

\newcommand{\vs}{\emph{vs. }}

\title{Overview of the TREC 2020 deep learning track}

\author[1]{Nick Craswell}
\author[1,2]{Bhaskar Mitra}
\author[2]{Emine Yilmaz}
\author[3]{Daniel Campos}
\affil[1]{Microsoft AI \& Research\\\texttt{\small \{nickcr, bmitra\}@microsoft.com}}
\affil[2]{University College London\\\texttt{\small \{bhaskar.mitra.15,emine.yilmaz\}@ucl.ac.uk}}
\affil[3]{University of Illinois Urbana-Champaign\\\texttt{\small \{dcampos3\}@illinois.edu}}

\begin{document}
\maketitle

\begin{abstract}

This is the second year of the TREC Deep Learning Track, with the goal of studying ad hoc ranking in the large training data regime. We again have a document retrieval task and a passage retrieval task, each with hundreds of thousands of human-labeled training queries. We evaluate using single-shot TREC-style evaluation, to give us a picture of which ranking methods work best when large data is available, with much more comprehensive relevance labeling on the small number of test queries. This year we have further evidence that rankers with BERT-style pretraining outperform other rankers in the large data regime.

\end{abstract}
\section{Introduction}
\label{sec:intro}

Deep learning methods, where a computational model learns an intricate representation of a large-scale dataset, yielded dramatic performance improvements in speech recognition and computer vision~\citep{lecun2015deep}. When we have seen such improvements, a common factor is the availability of large-scale training data~\citep{deng2009imagenet, bellemare2013arcade}. For ad hoc ranking in information retrieval, which is a core problem in the field, we did not initially see dramatic improvements in performance from deep learning methods. This led to questions about whether deep learning methods were helping at all~\citep{Yang2019critically}. If large training data sets are a factor, one explanation for this could be that the training sets were too small.

The TREC Deep Learning Track, and associated MS MARCO leaderboards \citep{bajaj2016ms}, have introduced human-labeled training sets that were previously unavailable. The main goal is to study information retrieval in the \emph{large training data} regime, to see which retrieval methods work best. 

The two tasks, document retrieval and passage retrieval, each have hundreds of thousands of human-labeled training queries. The training labels are sparse, with often only one positive example per query. Unlike the MS MARCO leaderboards, which evaluate using the same kind of sparse labels, the evaluation at TREC uses much more comprehensive relevance labeling. Each year of TREC evaluation evaluates on a new set of test queries, where participants submit before the test labels have even been generated, so the TREC results are the gold standard for avoiding multiple testing and overfitting. However, the comprehensive relevance labeling also generates a reusable test collections, allowing reuse of the dataset in future studies, although people should be careful to avoid overfitting and overiteration. 

The main goals of the Deep Learning Track in 2020 have been:
\begin{enumerate*}[label=\arabic*)]
\item To provide large reusable training datasets with associated large scale click dataset for training deep learning and traditional ranking methods in a large training data regime, 
\item To construct reusable test collections for evaluating quality of deep learning and traditional ranking methods,
\item To perform a rigorous blind single-shot evaluation, where test labels don't even exist until after all runs are submitted, to compare different ranking methods, and
\item To study this in both a traditional TREC setup with end-to-end retrieval and in a re-ranking setup that matches how some models may be deployed in practice.
\end{enumerate*}

\section{Task description}
\label{sec:task}

The track has two tasks: Document retrieval and passage retrieval.
Participants were allowed to submit up to three runs per task, although this was not strictly enforced.
Submissions to both tasks used the same set of $200$ test queries. In the pooling and judging process, NIST chose a subset of the queries for judging, based on budget constraints and with the goal of finding a sufficiently comprehensive set of relevance judgments to make the test collection reusable. This led to a judged test set of $45$ queries for document retrieval and $54$ queries for passage retrieval. The document queries are not a subset of the passage queries.

When submitting each run, participants indicated what external data, pretrained models and other resources were used, as well as information on what style of model was used.
Below we provide more detailed information about the document retrieval and passage retrieval tasks, as well as the datasets provided as part of these tasks. 

\subsection{Document retrieval task}

The first task focuses on document retrieval, with two subtasks:
\begin{enumerate*}[label=(\roman*)]
    \item Full retrieval and
    \item top-$100$ reranking.
\end{enumerate*}

In the full retrieval subtask, the runs are expected to rank documents based on their relevance to the query, where documents can be retrieved from the full document collection provided. This subtask models the end-to-end retrieval scenario.

In the reranking subtask, participants were provided with an initial ranking of $100$ documents, giving all participants the same starting point. This is a common scenario in many real-world retrieval systems that employ a telescoping architecture \citep{matveeva2006high, wang2011cascade}. The reranking subtask allows participants to focus on learning an effective relevance estimator, without the need for implementing an end-to-end retrieval system. It also makes the reranking runs more comparable, because they all rerank the same set of 100 candidates.

The initial top-$100$ rankings were retrieved using Indri \citep{strohman2005indri} on the full corpus with Krovetz stemming and stopwords eliminated. 



Judgments are on a four-point scale:
\begin{etaremune}[start=3]
    \item \textbf{Perfectly relevant:} Document is dedicated to the query, it is worthy of being a top result in a search engine.
    \item \textbf{Highly relevant:} The content of this document provides substantial information on the query.
    \item \textbf{Relevant:} Document provides some information relevant to the query, which may be minimal.
    \item \textbf{Irrelevant:} Document does not provide any useful information about the query.
\end{etaremune}
For metrics that binarize the judgment scale, we map document judgment levels 3,2,1 to relevant and map document judgment level 0 to irrelevant.

\subsection{Passage retrieval task}
Similar to the document retrieval task, the passage retrieval task includes
\begin{enumerate*}[label=(\roman*)]
    \item a full retrieval and
    \item a top-$1000$ reranking tasks.
\end{enumerate*}

In the full retrieval subtask, given a query, the participants were expected to retrieve a ranked list of passages from the full collection based on their estimated likelihood of containing an answer to the question.
Participants could submit up to $1000$ passages per query for this end-to-end retrieval task.

In the top-$1000$ reranking subtask, $1000$ passages per query were provided to participants, giving all participants the same starting point. The sets of 1000 were generated based on BM25 retrieval with no stemming as applied to the full collection. Participants were expected to rerank the 1000 passages based on their estimated likelihood of containing an answer to the query.
In this subtask, we can compare different reranking methods based on the same initial set of $1000$ candidates, with the same rationale as described for the document reranking subtask.

Judgments are on a four-point scale:
\begin{etaremune}[start=3]
    \item \textbf{Perfectly relevant:} The passage is dedicated to the query and contains the exact answer.
    \item \textbf{Highly relevant:} The passage has some answer for the query, but the answer may be a bit unclear, or hidden amongst extraneous information.
    \item \textbf{Related:} The passage seems related to the query but does not answer it.
    \item \textbf{Irrelevant:} The passage has nothing to do with the query.
\end{etaremune}
For metrics that binarize the judgment scale, we map passage judgment levels 3,2 to relevant and map document judgment levels 1,0 to irrelevant.



\section{Datasets}
\label{sec:data}

Both tasks have large training sets based on human relevance assessments, derived from MS MARCO.
These are sparse, with no negative labels and often only one positive label per query, analogous to some real-world training data such as click logs.

In the case of passage retrieval, the positive label indicates that the passage contains an answer to a query. 
In the case of document retrieval, we transferred the passage-level label to the corresponding source document that contained the passage. We do this under the assumption that a document with a relevant passage is a relevant document, although we note that our document snapshot was generated at a different time from the passage dataset, so there can be some mismatch. Despite this, machine learning models trained with these labels seem to benefit from using the labels, when evaluated using NIST's non-sparse, non-transferred labels. This suggests the transferred document labels are meaningful for our TREC task.

This year for the document retrieval task, we also release a large scale click dataset, The ORCAS data, constructed from the logs of a major search engine~\citep{craswell2020orcas}.  The data could be used in a variety of ways, for example as additional training data (almost 50 times larger than the main training set) or as a document field in addition to title, URL and body text fields available in the original training data.

For each task there is a corresponding MS MARCO leaderboard, using the same corpus and sparse training data, but using sparse data for evaluation as well, instead of the NIST test sets. We analyze the agreement between the two types of test in Section~\ref{sec:result}.


\begin{table}[]
    \centering
    \caption{Summary of statistics on TREC 2020 Deep Learning Track datasets.}

\begin{tabular}{@{}lrr@{}}
\toprule
~                     & \multicolumn{1}{c}{Document task} & \multicolumn{1}{c}{Passage task} \\ 
Data                  & \multicolumn{1}{c}{Number of records} & \multicolumn{1}{c}{Number of records} \\ 
\midrule
Corpus                & $3,213,835$    & $8,841,823$       \\
\addlinespace
Train queries         & $367,013$      & $502,939$         \\
Train qrels           & $384,597$      & $532,761$         \\
\addlinespace
Dev queries           & $5,193$        & $6,980$          \\
Dev qrels             & $5,478$        & $7,437$          \\
\addlinespace
2019 TREC queries    & $200 \rightarrow 43$          & $200 \rightarrow 43$             \\
2019 TREC qrels      & $16,258$       & $9,260$          \\
\addlinespace
2020 TREC queries          & $200 \rightarrow 45$          & $200 \rightarrow 54$             \\
2020 TREC qrels            & $9,098$    &  $11,386 $        \\
\bottomrule
\end{tabular}

    \label{tbl:data}
\end{table}

\begin{table*}
\centering
\caption{Summary of ORCAS data. Each record in the main file (\texttt{orcas.tsv}) indicates a click between a query (Q) and a URL (U), also listing a query ID (QID) and the corresponding TREC document ID (DID). The run file is the top-100 using Indri query likelihood, for use as negative samples during training.}
\begin{tabular}{lrrl}
\toprule
Filename & Number of records & Data in each record \\
\midrule
\texttt{orcas.tsv} & 18.8M & \texttt{QID Q DID U}\\
\texttt{orcas-doctrain-qrels.tsv} & 18.8M & \texttt{QID DID} \\
\texttt{orcas-doctrain-queries.tsv} & 10.4M & \texttt{QID Q}\\ 
\texttt{orcas-doctrain-top100} & 983M & \texttt{QID DID score} \\

\bottomrule
\end{tabular}
\label{tab:orcas}
\end{table*}

Table \ref{tbl:data} and Table \ref{tab:orcas} provide descriptive statistics for the dataset derived from MS MARCO and the ORCAS dataset, respectively.
More details about the datasets---including directions for download---is available on the TREC 2020 Deep Learning Track website\footnote{\url{https://microsoft.github.io/TREC-2020-Deep-Learning}}.
Interested readers are also encouraged to refer to \citep{bajaj2016ms} for details on the original MS MARCO dataset.
\begin{table}
    \centering
    \caption{Summary of statistics of runs for the two retrieval tasks at the TREC 2020 Deep Learning Track.}
    \begin{tabular}{lrr}
    \hline
    \hline
        & \textbf{Document retrieval} & \textbf{Passage retrieval} \\
        \hline
        Number of groups & 14 & 14 \\
        Number of total runs & 64 & 59 \\
        Number of runs w/ category: nnlm & 27 & 43 \\
        Number of runs w/ category: nn & 11 & 2 \\
        Number of runs w/ category: trad & 26 & 14 \\
        Number of runs w/ category: rerank & 19 & 18 \\
        Number of runs w/ category: fullrank & 45 & 41 \\
        \hline
        \hline
    \end{tabular}
    \label{tbl:runs-by-type}
\end{table}

\begin{figure}
  \center
  \begin{subfigure}{.49\textwidth}
    \includegraphics[width=\textwidth]{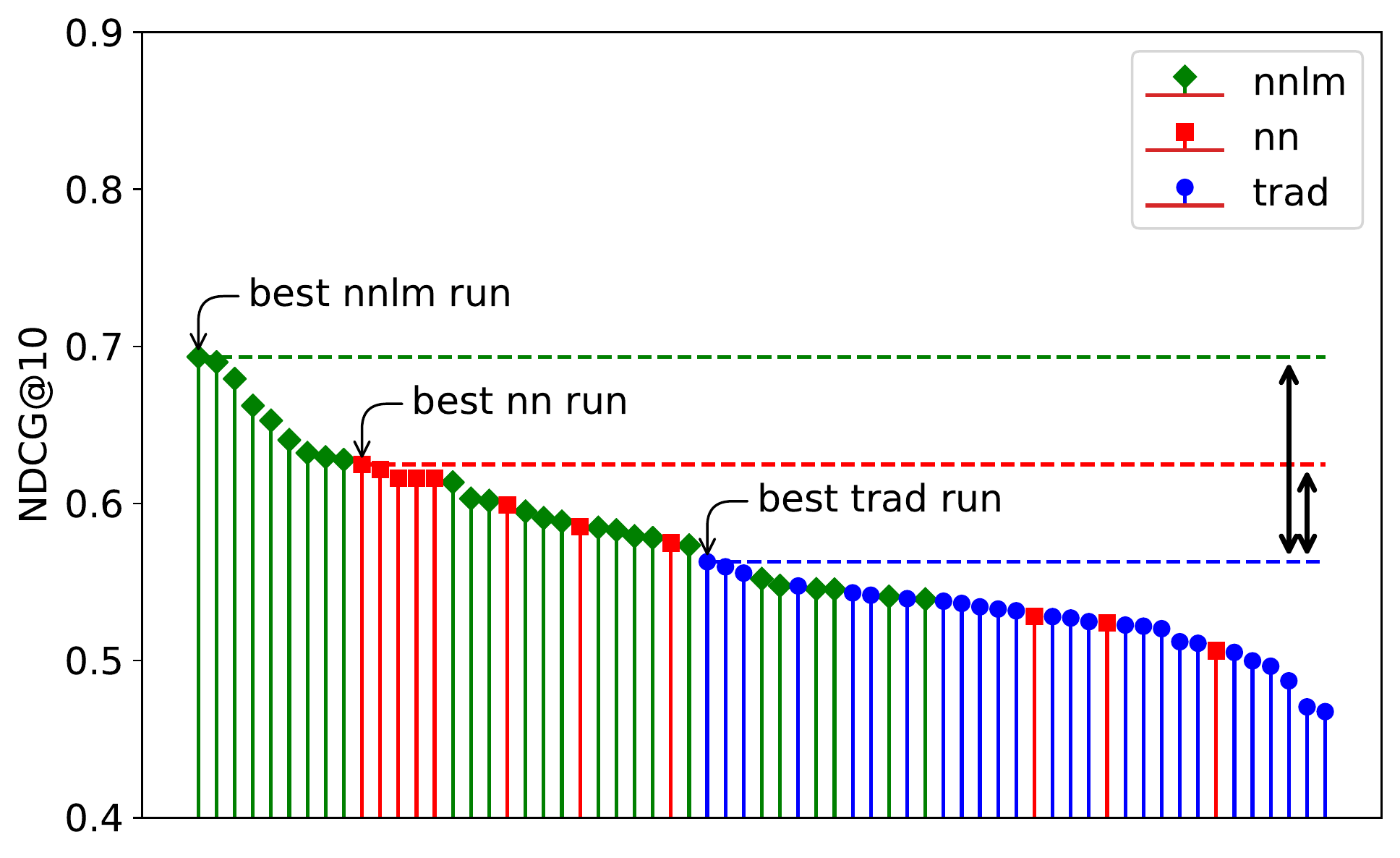}
    \caption{Document retrieval task}
    \label{fig:model-task-docs-stem-by-model-type}
  \end{subfigure}
  \hfill
  \begin{subfigure}{.49\textwidth}
    \includegraphics[width=\textwidth]{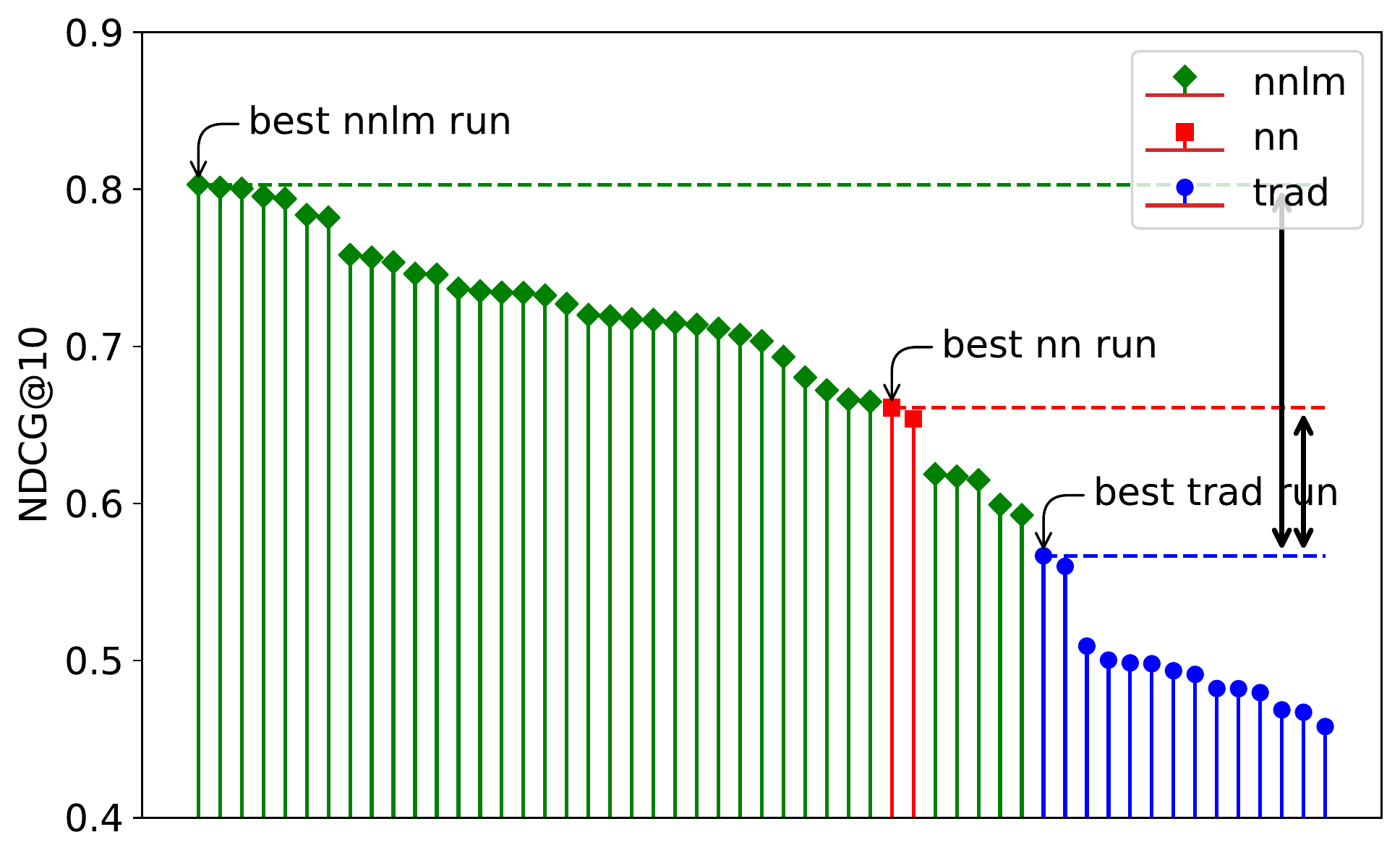}
    \caption{Passage retrieval task}
    \label{fig:model-task-passages-stem-by-model-type}
  \end{subfigure}
  \caption{NDCG@10 results, broken down by run type. Runs of type ``nnlm'', meaning they use language models such as BERT, performed best on both tasks. Other neural network models ``nn'' and non-neural models ``trad'' had relatively lower performance this year. More iterations of evaluation and analysis would be needed to determine if this is a general result, but it is a strong start for the argument that deep learning methods may take over from traditional methods in IR applications.}
  \label{fig:model-stem-by-model-type}
\end{figure}

\section{Results and analysis}
\label{sec:result}

\paragraph{Submitted runs}
The TREC 2020 Deep Learning Track had 25 participating groups, with a total of 123 runs submitted across both tasks.

Based run submission surveys, we manually classify each run into one of three categories:
\begin{itemize}
    \item \textbf{nnlm:} if the run employs large scale pre-trained neural language models, such as BERT \citep{devlin2018bert} or XLNet \citep{yang2019xlnet}
    \item \textbf{nn:} if the run employs some form of neural network based approach---\eg, Duet \citep{mitra2017learning, mitra2019updated} or using word embeddings \citep{joulin2016bag}---but does not fall into the ``nnlm'' category
    \item \textbf{trad:} if the run exclusively uses traditional IR methods like BM25 \citep{robertson2009probabilistic} and RM3 \citep{abdul2004umass}.
\end{itemize}

We placed 70 ($57\%$) runs in the ``nnlm'' category, 13 ($10\%$) in the ``nn'' category, and the remaining 40 ($33\%$) in the ``trad'' category. In 2019, 33 ($44\%$) runs were in the ``nnlm'' category, 20 ($27\%$) in the ``nn'' category, and the remaining 22 ($29\%$) in the ``trad'' category. While there was a significant increase in the total number of runs submitted compared to last year, we observed a significant reduction in the fraction of runs in the ``nn'' category.

We further categorize runs based on subtask:
\begin{itemize}
    \item \textbf{rerank:} if the run reranks the provided top-$k$ candidates, or
    \item \textbf{fullrank:} if the run employs their own phase 1 retrieval system.
\end{itemize}

We find that only 37 ($30\%$) submissions fall under the ``rerank'' category---while the remaining 86 ($70\%$) are ``fullrank''.
Table~\ref{tbl:runs-by-type} breaks down the submissions by category and task.

\paragraph{Overall results}

Our main metric in both tasks is Normalized Discounted Cumulative Gain (NDCG)---specifically, NDCG@10, since it makes use of our 4-level judgments and focuses on the first results that users will see. To get a picture of the ranking quality outside the top-10 we also report Average Precision (AP), although this binarizes the judgments. For comparison to the MS MARCO leaderboard, which often only has one relevant judgment per query, we report the Reciprocal Rank (RR) of the first relevant document on the NIST judgments, and also using the sparse leaderboard judgments.

Some of our evaluation is concerned with the quality of the top-$k$ results, where $k=100$ for the document task and $k=1000$ for the passage task. We want to consider the quality of the top-$k$ set without considering how they are ranked, so we can see whether improving the set-based quality is correlated with an improvement in NDCG@10. Although we could use Recall@$k$ as a metric here, it binarizes the judgments, so we instead use Normalized Cumulative Gain (NCG@$k$)~\citep{rosset2018optimizing}. NCG is not supported in trec\_eval. For trec\_eval metrics that are correlated, see Recall@$k$ and NDCG@$k$.

The overall results are presented in Table~\ref{tbl:results-docs} for document retrieval and Table~\ref{tbl:results-passages} for passage retrieval. These tables include multiple metrics and run categories, which we now use in our analysis.

\paragraph{Neural \vs traditional methods.}

The first question we investigated as part of the track is which ranking methods work best in the large-data regime. We summarize NDCG@10 results by run type in Figure~\ref{fig:model-stem-by-model-type}.

For document retrieval runs (Figure~\ref{fig:model-task-docs-stem-by-model-type}) the best ``trad'' run is outperformed by ``nn'' and ``nnlm'' runs by several percentage points, with ``nnlm'' also having an advantage over ``nn''. We saw a similar pattern in our 2019 results. This year we encouraged submission of a variety of ``trad'' runs from different participating groups, to give ``trad'' more chances to outperform other run types. The best performing run of each category is indicated, with the best ``nnlm'' and ``nn'' models outperforming the best ``trad'' model by $23\%$ and $11\%$ respectively.

For passage retrieval runs (Figure~\ref{fig:model-task-passages-stem-by-model-type}) the gap between the best ``nnlm'' and ``nn'' runs and the best ``trad'' run is larger, at $42\%$ and $17\%$ respectively. One explanation for this could be that vocabulary mismatch between queries and relevant results is greater in short text, so neural methods that can overcome such mismatch have a relatively greater advantage in passage retrieval. Another explanation could be that there is already a public leaderboard, albeit without test labels from NIST, for the passage task. (We did not launch the document ranking leaderboard until after our 2020 TREC submission deadline.) In passage ranking, some TREC participants may have submitted neural models multiple times to the public leaderboard, so are relatively more experienced working with the passage dataset than the document dataset.

\begin{figure}
\includegraphics[width=\textwidth]{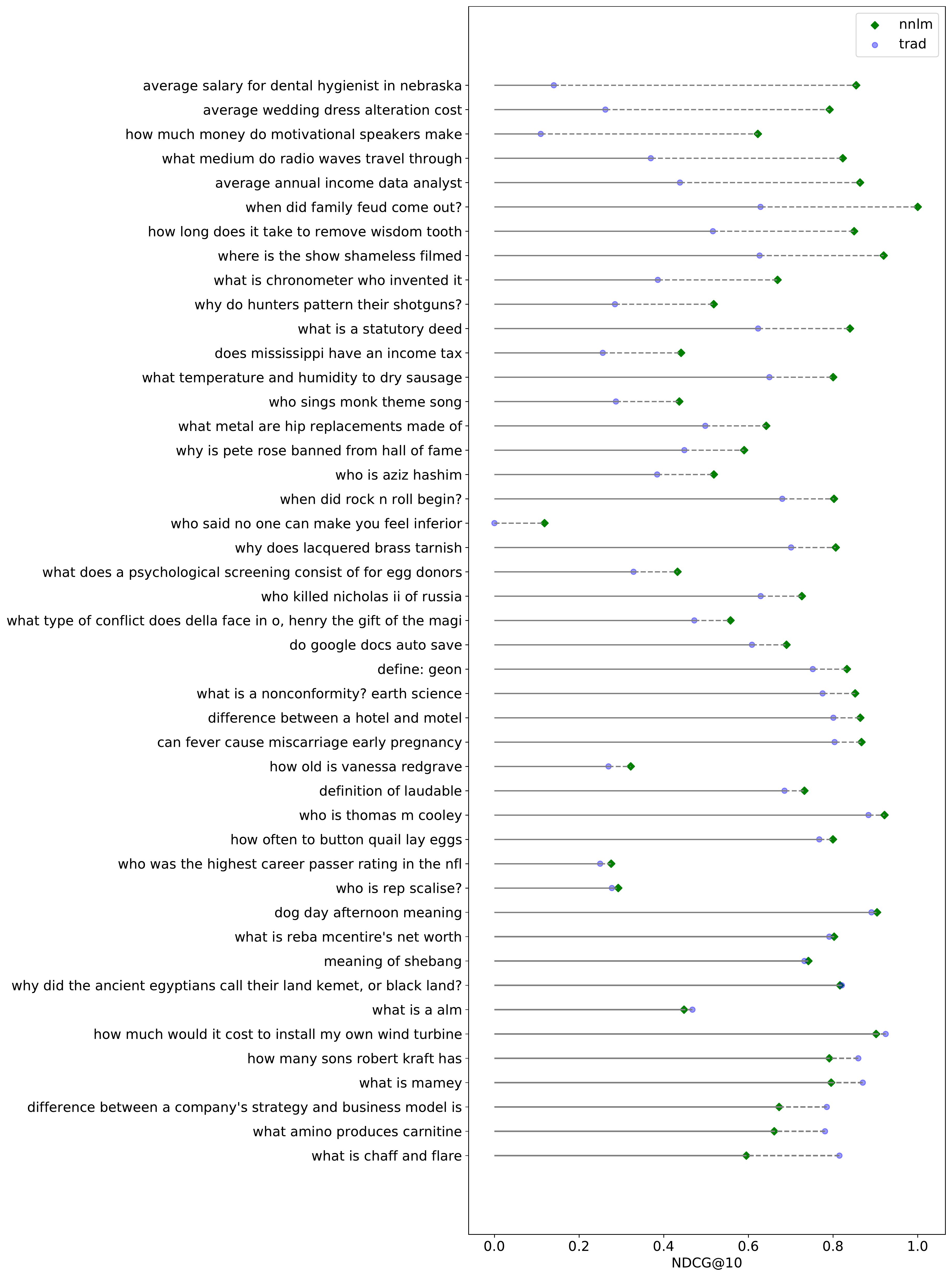}
\caption{Comparison of the best ``nnlm'' and ``trad'' runs on individual test queries for the document retrieval task. Queries are sorted by difference in mean performance between ``nnlm'' and ``trad'' runs. Queries on which ``nnlm'' wins with large margin are at the top.}
\label{fig:model-task-docs-bar-per-query}
\end{figure}

\begin{figure}
\includegraphics[width=\textwidth]{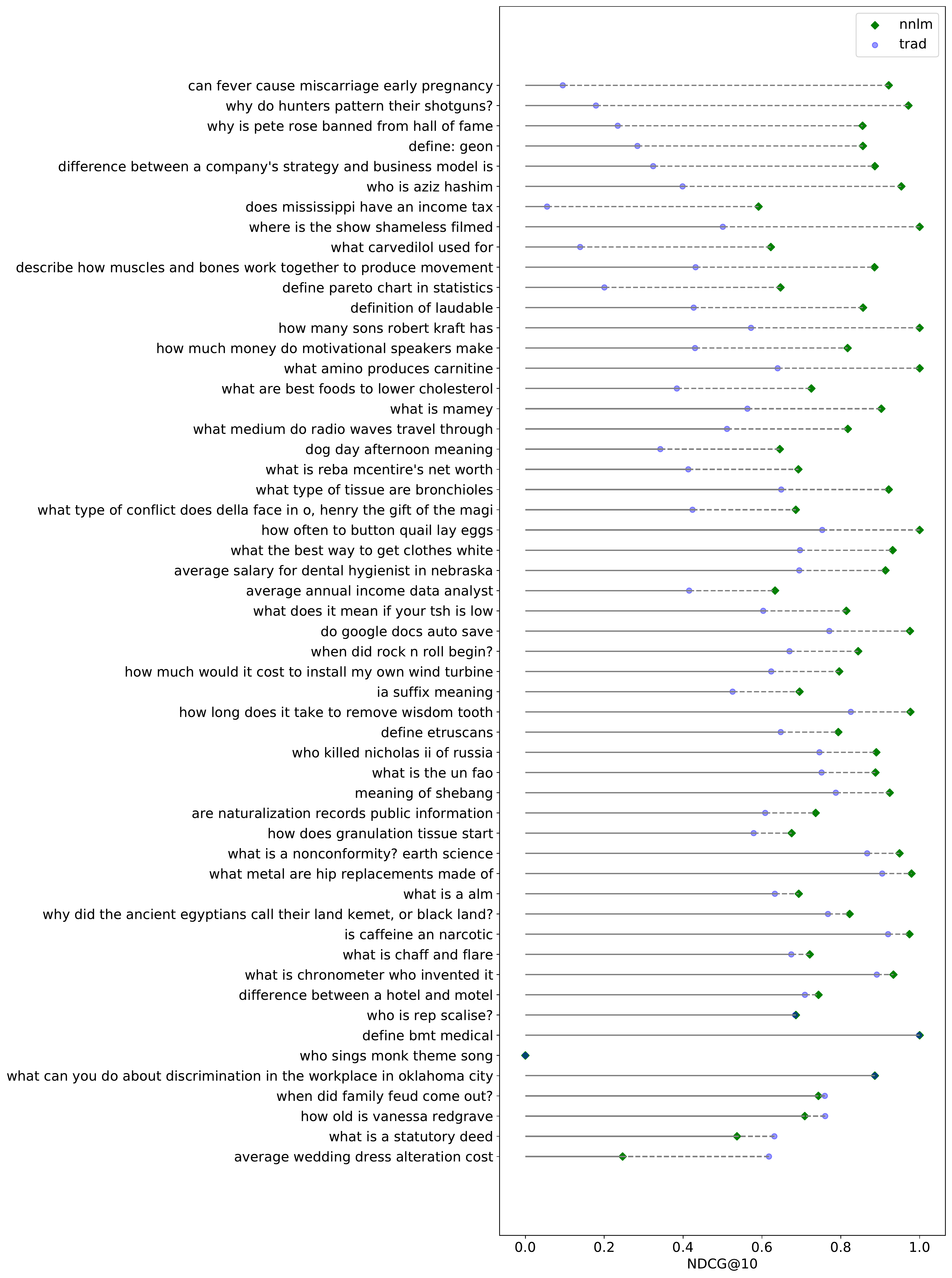}
\caption{Comparison of the best ``nnlm'' and ``trad'' runs on individual test queries for the passage retrieval task. Queries are sorted by difference in mean performance between ``nnlm'' and ``trad'' runs. Queries on which ``nnlm'' wins with large margin are at the top.}
\label{fig:model-task-passages-bar-per-query}
\end{figure}

In query-level win-loss analysis for the document retrieval task (Figure~\ref{fig:model-task-docs-bar-per-query}) the best ``nnlm'' model outperforms the best ``trad'' run on 38 out of the 45 test queries (\ie, $84\%$). Passage retrieval shows a similar pattern in Figure~\ref{fig:model-task-passages-bar-per-query}. Similar to last year's data, neither task has a large class of queries where the “nnlm” model performs worse. 

\begin{table}
    \fontsize{8}{9}\selectfont
    \centering
    \caption{Document retrieval runs. RR (MS) is based on MS MARCO labels. All other metrics are based on NIST labels. Rows are sorted by NDCG@10.}
\begin{tabular}{llllrrrrr}
\toprule
run & group & subtask & neural &  RR (MS) & RR &  NDCG@10 & NCG@100 & AP \\
\midrule
d\_d2q\_duo & h2oloo & fullrank & nnlm & 0.4451 & 0.9476 & 0.6934 & 0.7718 & 0.5422 \\
d\_d2q\_rm3\_duo & h2oloo & fullrank & nnlm & 0.4541 & 0.9476 & 0.6900 & 0.7769 & 0.5427 \\
d\_rm3\_duo & h2oloo & fullrank & nnlm & 0.4547 & 0.9476 & 0.6794 & 0.7498 & 0.5270 \\
ICIP\_run1 & ICIP & rerank & nnlm & 0.3898 & 0.9630 & 0.6623 & 0.6283 & 0.4333 \\
ICIP\_run3 & ICIP & rerank & nnlm & 0.4479 & 0.9667 & 0.6528 & 0.6283 & 0.4360 \\
fr\_doc\_roberta & BITEM & fullrank & nnlm & 0.3943 & 0.9365 & 0.6404 & 0.6806 & 0.4423 \\
ICIP\_run2 & ICIP & rerank & nnlm & 0.4081 & 0.9407 & 0.6322 & 0.6283 & 0.4206 \\
roberta-large & BITEM & rerank & nnlm & 0.3782 & 0.9185 & 0.6295 & 0.6283 & 0.4199 \\
bcai\_bertb\_docv & bcai & fullrank & nnlm & 0.4102 & 0.9259 & 0.6278 & 0.6604 & 0.4308 \\
ndrm3-orc-full & MSAI & fullrank & nn & 0.4369 & 0.9444 & 0.6249 & 0.6764 & 0.4280 \\
ndrm3-orc-re & MSAI & rerank & nn & 0.4451 & 0.9241 & 0.6217 & 0.6283 & 0.4194 \\
ndrm3-full & MSAI & fullrank & nn & 0.4213 & 0.9333 & 0.6162 & 0.6626 & 0.4069 \\
ndrm3-re & MSAI & rerank & nn & 0.4258 & 0.9333 & 0.6162 & 0.6283 & 0.4122 \\
ndrm1-re & MSAI & rerank & nn & 0.4427 & 0.9333 & 0.6161 & 0.6283 & 0.4150 \\
mpii\_run2 & mpii & rerank & nnlm & 0.3228 & 0.8833 & 0.6135 & 0.6283 & 0.4205 \\
bigIR-DTH-T5-R & QU & rerank & nnlm & 0.3235 & 0.9119 & 0.6031 & 0.6283 & 0.3936 \\
mpii\_run1 & mpii & rerank & nnlm & 0.3503 & 0.9000 & 0.6017 & 0.6283 & 0.4030 \\
ndrm1-full & MSAI & fullrank & nn & 0.4350 & 0.9333 & 0.5991 & 0.6280 & 0.3858 \\
uob\_runid3 & UoB & rerank & nnlm & 0.3294 & 0.9259 & 0.5949 & 0.6283 & 0.3948 \\
bigIR-DTH-T5-F & QU & fullrank & nnlm & 0.3184 & 0.8916 & 0.5907 & 0.6669 & 0.4259 \\
d\_d2q\_bm25 & anserini & fullrank & nnlm & 0.3338 & 0.9369 & 0.5885 & 0.6752 & 0.4230 \\
TUW-TKL-2k & TU\_Vienna & rerank & nn & 0.3683 & 0.9296 & 0.5852 & 0.6283 & 0.3810 \\
bigIR-DH-T5-R & QU & rerank & nnlm & 0.2877 & 0.8889 & 0.5846 & 0.6283 & 0.3842 \\
uob\_runid2 & UoB & rerank & nnlm & 0.3534 & 0.9100 & 0.5830 & 0.6283 & 0.3976 \\
uogTrQCBMP & UoGTr & fullrank & nnlm & 0.3521 & 0.8722 & 0.5791 & 0.6034 & 0.3752 \\
uob\_runid1 & UoB & rerank & nnlm & 0.3124 & 0.8852 & 0.5781 & 0.6283 & 0.3786 \\
TUW-TKL-4k & TU\_Vienna & rerank & nn & 0.4097 & 0.9185 & 0.5749 & 0.6283 & 0.3749 \\
bigIR-DH-T5-F & QU & fullrank & nnlm & 0.2704 & 0.8902 & 0.5734 & 0.6669 & 0.4177 \\
bl\_bcai\_multfld & bl\_bcai & fullrank & trad & 0.2622 & 0.9195 & 0.5629 & 0.6299 & 0.3829 \\
indri-sdmf & RMIT & fullrank & trad & 0.3431 & 0.8796 & 0.5597 & 0.6908 & 0.3974 \\
bcai\_classic & bcai & fullrank & trad & 0.3082 & 0.8648 & 0.5557 & 0.6420 & 0.3906 \\
longformer\_1 & USI & rerank & nnlm & 0.3614 & 0.8889 & 0.5520 & 0.6283 & 0.3503 \\
uogTr31oR & UoGTr & fullrank & nnlm & 0.3257 & 0.8926 & 0.5476 & 0.5496 & 0.3468 \\
rterrier-expC2 & bl\_rmit & fullrank & trad & 0.3122 & 0.8259 & 0.5475 & 0.6442 & 0.3805 \\
bigIR-DT-T5-R & QU & rerank & nnlm & 0.2293 & 0.9407 & 0.5455 & 0.6283 & 0.3373 \\
uogTrT20 & UoGTr & fullrank & nnlm & 0.3787 & 0.8711 & 0.5453 & 0.5354 & 0.3692 \\
RMIT\_DFRee & RMIT & fullrank & trad & 0.2984 & 0.8756 & 0.5431 & 0.6979 & 0.4087 \\
rmit\_indri-fdm & bl\_rmit & fullrank & trad & 0.2779 & 0.8481 & 0.5416 & 0.6812 & 0.3859 \\
d\_d2q\_bm25rm3 & anserini & fullrank & nnlm & 0.2314 & 0.8147 & 0.5407 & 0.6831 & 0.4228 \\
rindri-bm25 & bl\_rmit & fullrank & trad & 0.3302 & 0.8572 & 0.5394 & 0.6503 & 0.3773 \\
bigIR-DT-T5-F & QU & fullrank & nnlm & 0.2349 & 0.9060 & 0.5390 & 0.6669 & 0.3619 \\
bl\_bcai\_model1 & bl\_bcai & fullrank & trad & 0.2901 & 0.8358 & 0.5378 & 0.6390 & 0.3774 \\
bl\_bcai\_prox & bl\_bcai & fullrank & trad & 0.2763 & 0.8164 & 0.5364 & 0.6405 & 0.3766 \\
terrier-jskls & bl\_rmit & fullrank & trad & 0.3190 & 0.8204 & 0.5342 & 0.6761 & 0.4008 \\
rmit\_indri-sdm & bl\_rmit & fullrank & trad & 0.2702 & 0.8470 & 0.5328 & 0.6733 & 0.3780 \\
rterrier-tfidf & bl\_rmit & fullrank & trad & 0.2869 & 0.8241 & 0.5317 & 0.6410 & 0.3734 \\
BIT-run2 & BIT.UA & fullrank & nn & 0.2687 & 0.8611 & 0.5283 & 0.6061 & 0.3466 \\
RMIT\_DPH & RMIT & fullrank & trad & 0.3117 & 0.8278 & 0.5280 & 0.6531 & 0.3879 \\
d\_bm25 & anserini & fullrank & trad & 0.2814 & 0.8521 & 0.5271 & 0.6453 & 0.3791 \\
d\_bm25rm3 & anserini & fullrank & trad & 0.2645 & 0.8541 & 0.5248 & 0.6632 & 0.4006 \\
BIT-run1 & BIT.UA & fullrank & nn & 0.3045 & 0.8389 & 0.5239 & 0.6061 & 0.3466 \\
rterrier-dph & bl\_rmit & fullrank & trad & 0.3033 & 0.8267 & 0.5226 & 0.6634 & 0.3884 \\
rterrier-tfidf2 & bl\_rmit & fullrank & trad & 0.3010 & 0.8407 & 0.5219 & 0.6287 & 0.3607 \\
uogTrBaseQL17o & bl\_uogTr & fullrank & trad & 0.4233 & 0.8276 & 0.5203 & 0.6028 & 0.3529 \\
uogTrBaseL17o & bl\_uogTr & fullrank & trad & 0.3870 & 0.7980 & 0.5120 & 0.5501 & 0.3248 \\
rterrier-dph\_sd & bl\_rmit & fullrank & trad & 0.3243 & 0.8296 & 0.5110 & 0.6650 & 0.3784 \\
BIT-run3 & BIT.UA & fullrank & nn & 0.2696 & 0.8296 & 0.5063 & 0.6072 & 0.3267 \\
uogTrBaseDPHQ & bl\_uogTr & fullrank & trad & 0.3459 & 0.8052 & 0.5052 & 0.6041 & 0.3461 \\
uogTrBaseQL16 & bl\_uogTr & fullrank & trad & 0.3321 & 0.7930 & 0.4998 & 0.6030 & 0.3436 \\
uogTrBaseL16 & bl\_uogTr & fullrank & trad & 0.3062 & 0.8219 & 0.4964 & 0.5495 & 0.3248 \\
uogTrBaseDPH & bl\_uogTr & fullrank & trad & 0.3179 & 0.8415 & 0.4871 & 0.5490 & 0.3070 \\
nlm-bm25-prf-2 & NLM & fullrank & trad & 0.2732 & 0.8099 & 0.4705 & 0.5218 & 0.2912 \\
nlm-bm25-prf-1 & NLM & fullrank & trad & 0.2390 & 0.8086 & 0.4675 & 0.4958 & 0.2720 \\
mpii\_run3 & mpii & rerank & nnlm & 0.1499 & 0.6388 & 0.3286 & 0.6283 & 0.2587 \\
\bottomrule
\end{tabular}
\label{tbl:results-docs}
\end{table}

\begin{table}
    \small
    \centering
    \caption{Passage retrieval runs. RR (MS) is based on MS MARCO labels. All other metrics are based on NIST labels.}
\begin{tabular}{llllrrrrr}
\toprule
run &         group &   subtask & neural &  RR (MS) &     RR &  NDCG@10 & NCG@1000 &   AP \\
\midrule
pash\_r3 & PASH & rerank & nnlm & 0.3678 & 0.9147 & 0.8031 & 0.7056 & 0.5445 \\
pash\_r2 & PASH & rerank & nnlm & 0.3677 & 0.9023 & 0.8011 & 0.7056 & 0.5420 \\
pash\_f3 & PASH & fullrank & nnlm & 0.3506 & 0.8885 & 0.8005 & 0.7255 & 0.5504 \\
pash\_f1 & PASH & fullrank & nnlm & 0.3598 & 0.8699 & 0.7956 & 0.7209 & 0.5455 \\
pash\_f2 & PASH & fullrank & nnlm & 0.3603 & 0.8931 & 0.7941 & 0.7132 & 0.5389 \\
p\_d2q\_bm25\_duo & h2oloo & fullrank & nnlm & 0.3838 & 0.8798 & 0.7837 & 0.8035 & 0.5609 \\
p\_d2q\_rm3\_duo & h2oloo & fullrank & nnlm & 0.3795 & 0.8798 & 0.7821 & 0.8446 & 0.5643 \\
p\_bm25rm3\_duo & h2oloo & fullrank & nnlm & 0.3814 & 0.8759 & 0.7583 & 0.7939 & 0.5355 \\
CoRT-electra & HSRM-LAVIS & fullrank & nnlm & 0.4039 & 0.8703 & 0.7566 & 0.8072 & 0.5399 \\
RMIT-Bart & RMIT & fullrank & nnlm & 0.3990 & 0.8447 & 0.7536 & 0.7682 & 0.5121 \\
pash\_r1 & PASH & rerank & nnlm & 0.3622 & 0.8675 & 0.7463 & 0.7056 & 0.4969 \\
NLE\_pr3 & NLE & fullrank & nnlm & 0.3691 & 0.8440 & 0.7458 & 0.8211 & 0.5245 \\
pinganNLP2 & pinganNLP & rerank & nnlm & 0.3579 & 0.8602 & 0.7368 & 0.7056 & 0.4881 \\
pinganNLP3 & pinganNLP & rerank & nnlm & 0.3653 & 0.8586 & 0.7352 & 0.7056 & 0.4918 \\
pinganNLP1 & pinganNLP & rerank & nnlm & 0.3553 & 0.8593 & 0.7343 & 0.7056 & 0.4896 \\
NLE\_pr2 & NLE & fullrank & nnlm & 0.3658 & 0.8454 & 0.7341 & 0.6938 & 0.5117 \\
NLE\_pr1 & NLE & fullrank & nnlm & 0.3634 & 0.8551 & 0.7325 & 0.6938 & 0.5050 \\
1 & nvidia\_ai\_apps & rerank & nnlm & 0.3709 & 0.8691 & 0.7271 & 0.7056 & 0.4899 \\
bigIR-BERT-R & QU & rerank & nnlm & 0.4040 & 0.8562 & 0.7201 & 0.7056 & 0.4845 \\
fr\_pass\_roberta & BITEM & fullrank & nnlm & 0.3580 & 0.8769 & 0.7192 & 0.7982 & 0.4990 \\
bigIR-DCT-T5-F & QU & fullrank & nnlm & 0.3540 & 0.8638 & 0.7173 & 0.8093 & 0.5004 \\
rr-pass-roberta & BITEM & rerank & nnlm & 0.3701 & 0.8635 & 0.7169 & 0.7056 & 0.4823 \\
bcai\_bertl\_pass & bcai & fullrank & nnlm & 0.3715 & 0.8453 & 0.7151 & 0.7990 & 0.4641 \\
bigIR-T5-R & QU & rerank & nnlm & 0.3574 & 0.8668 & 0.7138 & 0.7056 & 0.4784 \\
2 & nvidia\_ai\_apps & fullrank & nnlm & 0.3560 & 0.8507 & 0.7113 & 0.7447 & 0.4866 \\
bigIR-T5-BERT-F & QU & fullrank & nnlm & 0.3916 & 0.8478 & 0.7073 & 0.8393 & 0.5101 \\
bigIR-T5xp-T5-F & QU & fullrank & nnlm & 0.3420 & 0.8579 & 0.7034 & 0.8393 & 0.5001 \\
nlm-ens-bst-2 & NLM & fullrank & nnlm & 0.3542 & 0.8203 & 0.6934 & 0.7190 & 0.4598 \\
nlm-ens-bst-3 & NLM & fullrank & nnlm & 0.3195 & 0.8491 & 0.6803 & 0.7594 & 0.4526 \\
nlm-bert-rr & NLM & rerank & nnlm & 0.3699 & 0.7785 & 0.6721 & 0.7056 & 0.4341 \\
relemb\_mlm\_0\_2 & UAmsterdam & rerank & nnlm & 0.2856 & 0.7677 & 0.6662 & 0.7056 & 0.4350 \\
nlm-prfun-bert & NLM & fullrank & nnlm & 0.3445 & 0.8603 & 0.6648 & 0.6927 & 0.4265 \\
TUW-TK-Sparse & TU\_Vienna & rerank & nn & 0.3188 & 0.7970 & 0.6610 & 0.7056 & 0.4164 \\
TUW-TK-2Layer & TU\_Vienna & rerank & nn & 0.3075 & 0.7654 & 0.6539 & 0.7056 & 0.4179 \\
p\_d2q\_bm25 & anserini & fullrank & nnlm & 0.2757 & 0.7326 & 0.6187 & 0.8035 & 0.4074 \\
p\_d2q\_bm25rm3 & anserini & fullrank & nnlm & 0.2848 & 0.7424 & 0.6172 & 0.8391 & 0.4295 \\
bert\_6 & UAmsterdam & rerank & nnlm & 0.3240 & 0.7386 & 0.6149 & 0.7056 & 0.3760 \\
CoRT-bm25 & HSRM-LAVIS & fullrank & nnlm & 0.2201 & 0.8372 & 0.5992 & 0.8072 & 0.3611 \\
CoRT-standalone & HSRM-LAVIS & fullrank & nnlm & 0.2412 & 0.8112 & 0.5926 & 0.6002 & 0.3308 \\
bl\_bcai\_mdl1\_vt & bl\_bcai & fullrank & trad & 0.1854 & 0.7037 & 0.5667 & 0.7430 & 0.3380 \\
bcai\_class\_pass & bcai & fullrank & trad & 0.1999 & 0.7115 & 0.5600 & 0.7430 & 0.3374 \\
bl\_bcai\_mdl1\_vs & bl\_bcai & fullrank & trad & 0.1563 & 0.6277 & 0.5092 & 0.7430 & 0.3094 \\
indri-fdm & bl\_rmit & fullrank & trad & 0.1798 & 0.6498 & 0.5003 & 0.7778 & 0.2989 \\
terrier-InL2 & bl\_rmit & fullrank & trad & 0.1864 & 0.6436 & 0.4985 & 0.7649 & 0.3135 \\
terrier-BM25 & bl\_rmit & fullrank & trad & 0.1631 & 0.6186 & 0.4980 & 0.7572 & 0.3021 \\
DLH\_d\_5\_t\_25 & RMIT & fullrank & trad & 0.1454 & 0.5094 & 0.4935 & 0.8175 & 0.3199 \\
indri-lmds & bl\_rmit & fullrank & trad & 0.1250 & 0.5866 & 0.4912 & 0.7741 & 0.2961 \\
indri-sdm & bl\_rmit & fullrank & trad & 0.1600 & 0.6239 & 0.4822 & 0.7726 & 0.2870 \\
p\_bm25rm3 & anserini & fullrank & trad & 0.1495 & 0.6360 & 0.4821 & 0.7939 & 0.3019 \\
p\_bm25 & anserini & fullrank & trad & 0.1786 & 0.6585 & 0.4796 & 0.7428 & 0.2856 \\
bm25\_bert\_token & UAmsterdam & fullrank & trad & 0.1576 & 0.6409 & 0.4686 & 0.7169 & 0.2606 \\
terrier-DPH & bl\_rmit & fullrank & trad & 0.1420 & 0.5667 & 0.4671 & 0.7353 & 0.2758 \\
TF\_IDF\_d\_2\_t\_50 & RMIT & fullrank & trad & 0.1391 & 0.5317 & 0.4580 & 0.7722 & 0.2923 \\
small\_1k & reSearch2vec & rerank & nnlm & 0.0232 & 0.2785 & 0.2767 & 0.7056 & 0.2112 \\
med\_1k & reSearch2vec & rerank & nnlm & 0.0222 & 0.2720 & 0.2708 & 0.7056 & 0.2081 \\
DoRA\_Large\_1k & reSearch2vec & rerank & nnlm & 0.0208 & 0.2740 & 0.2661 & 0.7056 & 0.2072 \\
DoRA\_Small & reSearch2vec & fullrank & nnlm & 0.0000 & 0.1287 & 0.0484 & 0.0147 & 0.0088 \\
DoRA\_Med & reSearch2vec & fullrank & nnlm & 0.0000 & 0.1075 & 0.0431 & 0.0147 & 0.0087 \\
DoRA\_Large & reSearch2vec & fullrank & nnlm & 0.0000 & 0.1111 & 0.0414 & 0.0146 & 0.0079 \\
\bottomrule
\end{tabular}
\label{tbl:results-passages}
\end{table}

\paragraph{End-to-end retrieval \vs reranking.}

Our datasets include top-$k$ candidate result lists, with 100 candidates per query for document retrieval and 1000 candidates per query for passage retrieval. Runs that simply rerank the provided candidates are ``rerank'' runs, whereas runs that perform end-to-end retrieval against the corpus, with millions of potential results, are ``fullrank'' runs. We would expect that a ``fullrank'' run should be able to find a greater number of relevant candidates than we provided, achieving higher NCG@$k$. A multi-stage ``fullrank'' run should also be able to optimize the stages jointly, such that early stages produce candidates that later stages are good at handling.

\begin{figure}
  \center
  \begin{subfigure}{.49\textwidth}
    \includegraphics[width=\textwidth]{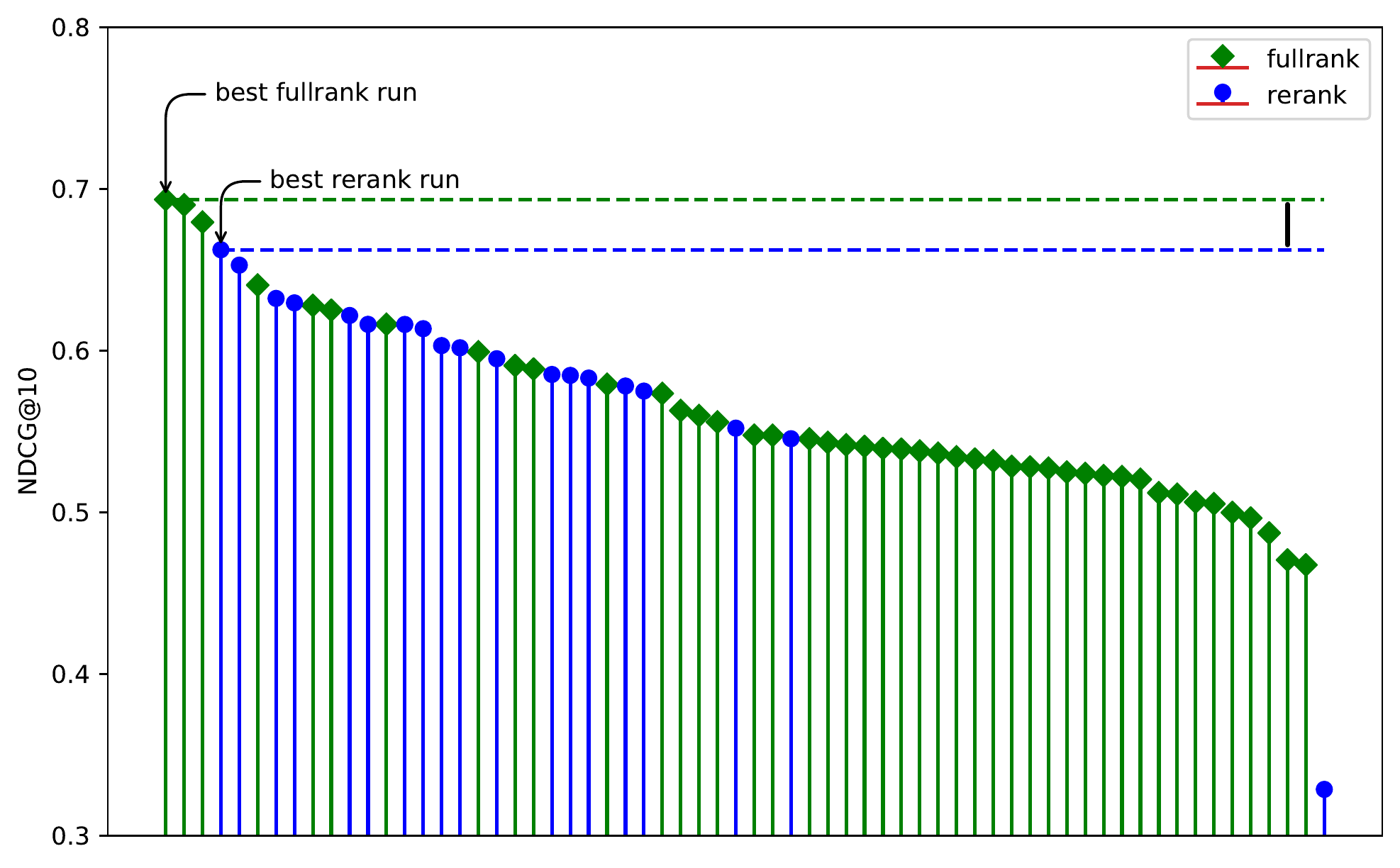}
    \caption{NDCG@10 for runs on the document retrieval task}
    \label{fig:model-task-docs-stem-by-subtask}
  \end{subfigure}
  \hfill
  \begin{subfigure}{.49\textwidth}
    \includegraphics[width=\textwidth]{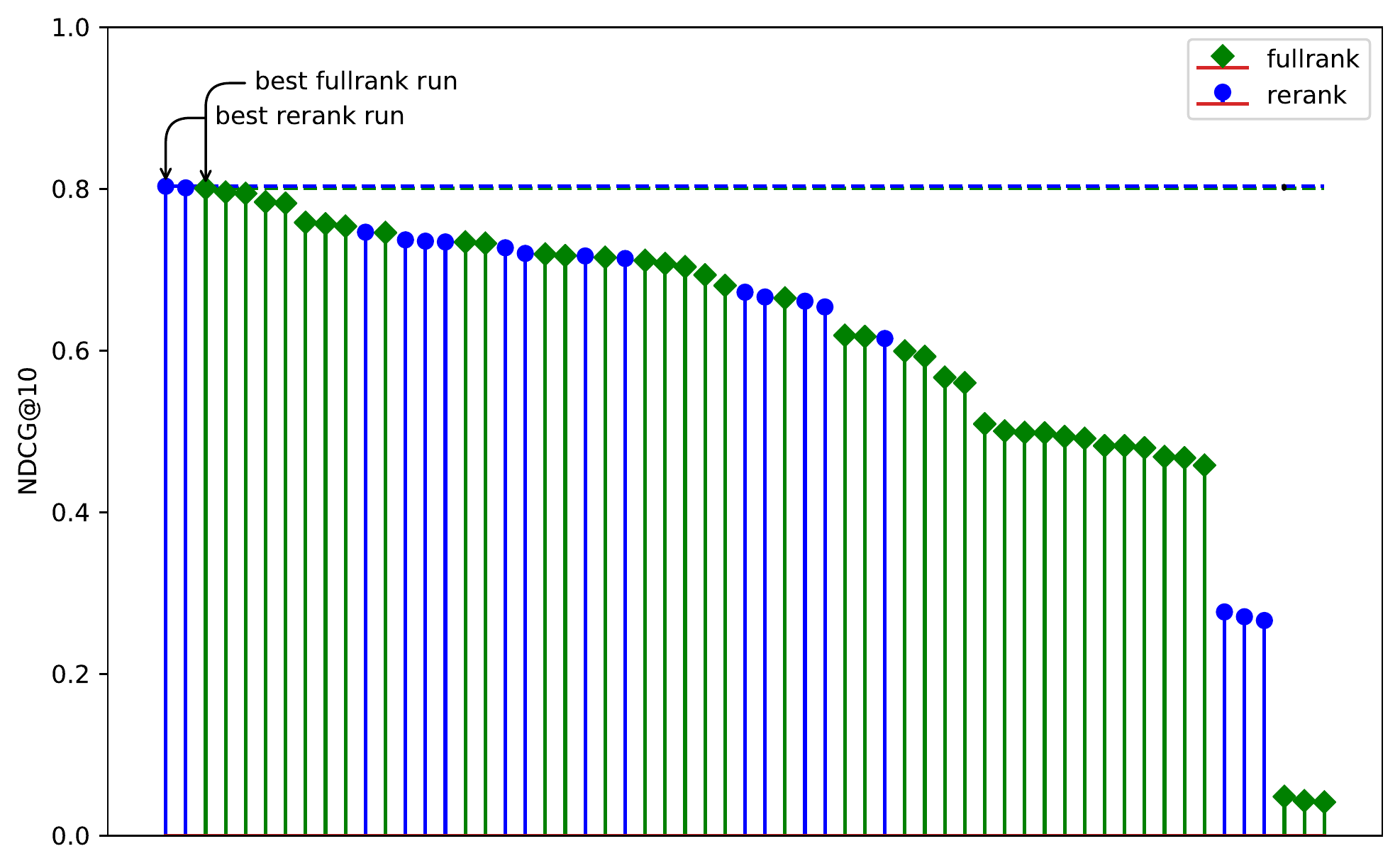}
    \caption{NDCG@10 for runs on the passage retrieval task}
    \label{fig:model-task-passages-stem-by-subtask}
  \end{subfigure}
  \begin{subfigure}{.49\textwidth}
    \includegraphics[width=\textwidth]{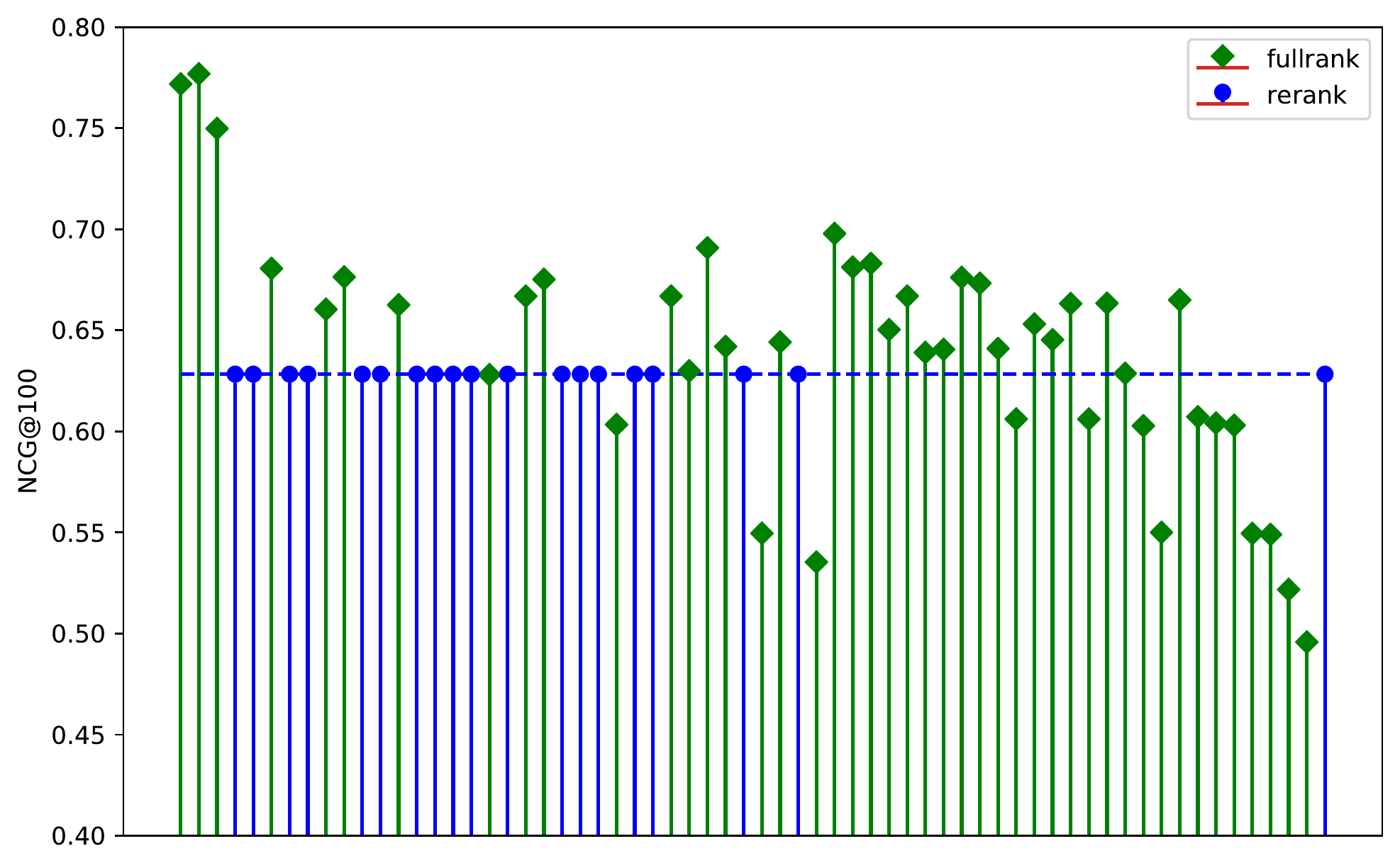}
    \caption{NCG@100 for runs on the document retrieval task}
    \label{fig:recall-task-docs-stem}
  \end{subfigure}
  \hfill
  \begin{subfigure}{.49\textwidth}
    \includegraphics[width=\textwidth]{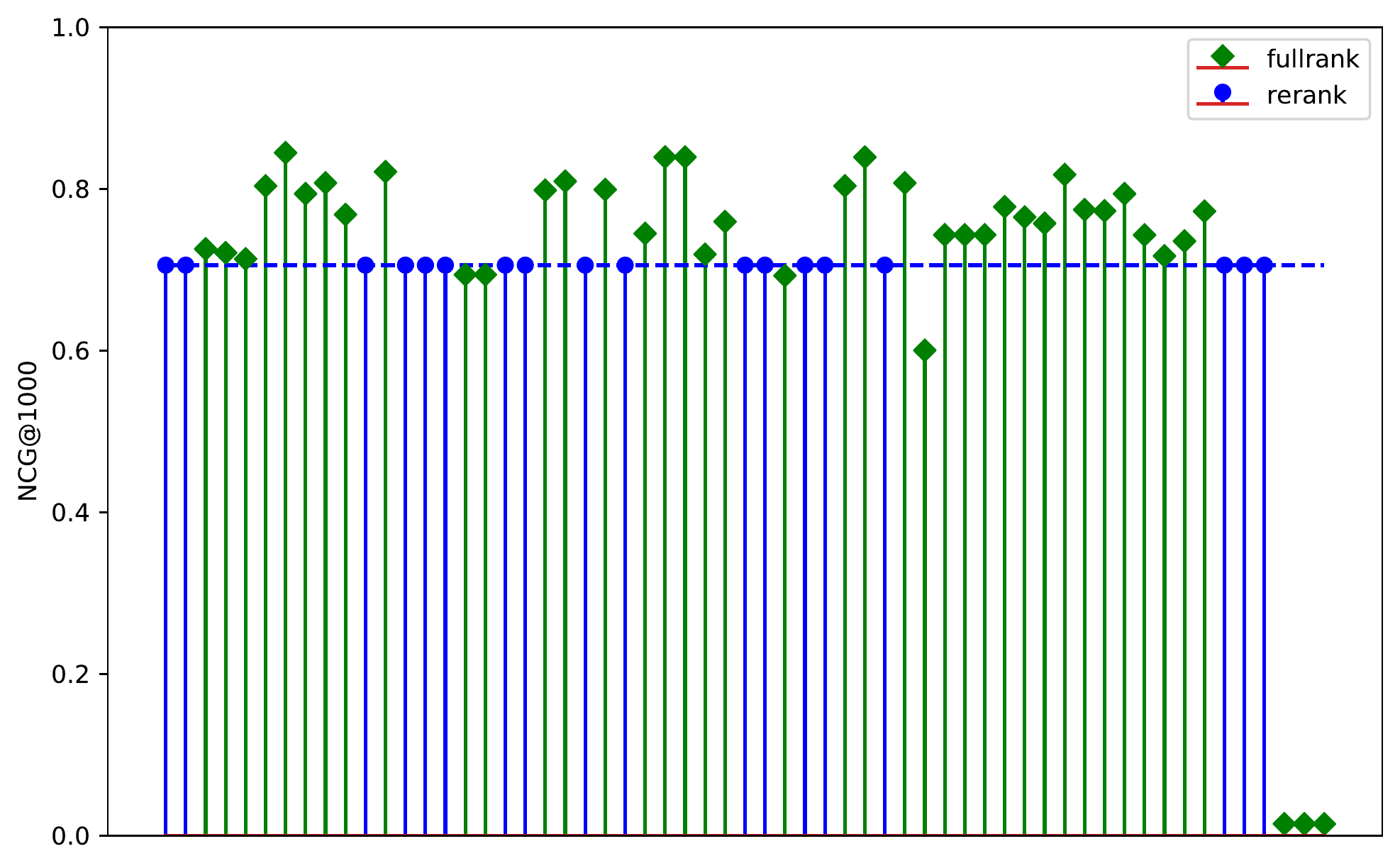}
    \caption{NCG@1000 for runs on the passage retrieval task}
    \label{fig:recall-task-passages-stem}
  \end{subfigure}
  \caption{Analyzing the impact of ``fullrank'' \vs ``rerank'' settings on retrieval performance.
  Figure~(a) and (b) show the performance of different runs on the document and passage retrieval tasks, respectively.
  Figure~(c) and (d) plot the NCG@100 and NCG@1000 metrics for the same runs for the two tasks, respectively.
  The runs are ordered by their NDCG@10 performance along the $x$-axis in all four plots.
  We observe, that the best run under the ``fullrank'' setting outperforms the same under the ``rerank'' setting for both document and passage retrieval tasks---although the gaps are relatively smaller compared to those in Figure~\ref{fig:model-stem-by-model-type}.
  If we compare Figure~(a) with (c) and Figure~(b) with (d), we do not observe any evidence that the NCG metric is a good predictor of NDCG@10 performance.}
  \label{fig:recall-stem}
\end{figure}

According to Figure~\ref{fig:recall-stem}, ``fullrank'' did not achieve much better NDCG@10 performance than ``rerank'' runs. In fact, for the passage retrieval task, the top two runs are of type ``rerank''. While it was possible for ``fullrank'' to achieve better NCG@$k$, it was also possible to make NCG@$k$ worse, and achieving significantly higher NCG@$k$ does not seem necessary to achieve good NDCG@10. 

Specifically, for the document retrieval task, the best ``fullrank'' run achieves $5\%$ higher NDCG@10 over the best ``rerank' run; whereas for the passage retrieval task, the best ``fullrank'' run performs slightly worse ($0.3\%$ lower NDCG@10) compared to the best ``rerank' run.

Similar to our observations from Deep Learning Track 2019, we are not yet seeing a strong advantage of ``fullrank'' over ``rerank''.
However, we hope that as the body of literature on neural methods for phase 1 retrieval (\eg, \citep{boytsov2016off, zamani2018neural, mitra2019incorporating, nogueira2019document}) grows, we would see a larger number of runs with deep learning as an ingredient for phase 1 in future editions of this TREC track.

\paragraph{Effect of ORCAS data}
Based on the descriptions provided, ORCAS data seems to have been used by six of the runs (ndrm3-orc-full, ndrm3-orc-re, uogTrBaseL17, uogTrBaseQL17o, uogTr31oR, relemb\_mlm\_0\_2). Most runs seem to be make use of the ORCAS data as a field, with some runs using the data as an additional training dataset as well. Most runs used the ORCAS data for the document retrieval task, with relemb\_mlm\_0\_2 being the only run using the ORCAS data for the passage retrieval task. 

This year it was not necessary to use ORCAS data to achieve the highest NDCG@10. However, when we compare the performance of the runs that use the ORCAS dataset with those that do not use the dataset within the same group, we observe that usage of the ORCAS dataset always led to an improved performance in terms of NDCG@10, with maximum increase being around $0.0513$ in terms of NDCG@10. This suggests that the ORCAS dataset is providing additional information that is not available in the training data. This could also imply that even though the training dataset provided as part of the track is very large, deep models are still in need of more training data.  

%

\paragraph{NIST labels \vs Sparse MS MARCO labels.}

Our baseline human labels from MS MARCO often have one known positive result per query. We use these labels for training, but they are also available for test queries. Although our official evaluation uses NDCG@10 with NIST labels, we now compare this with reciprocal rank (RR) using MS MARCO labels. Our goal is to understand how changing the labeling scheme and metric affects the overall results of the track, but if there is any disagreement we believe the NDCG results are more valid, since they evaluate the ranking more comprehensively and a ranker that can only perform well on labels with exactly the same distribution as the training set is not robust enough for use in real-world applications, where real users will have opinions that are not necessarily identical to the preferences encoded in sparse training labels.

\begin{figure}
    \centering
    \includegraphics[width=0.49\textwidth]{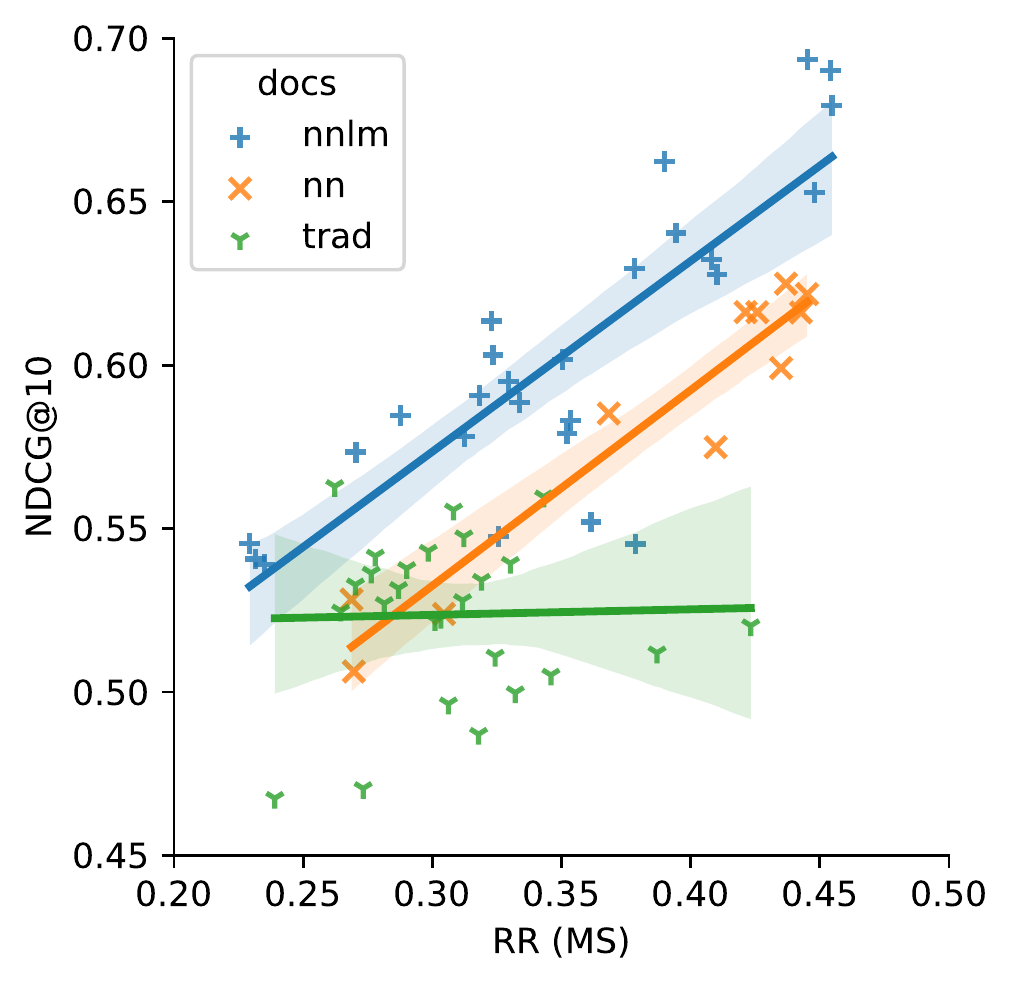}
    \includegraphics[width=0.49\textwidth]{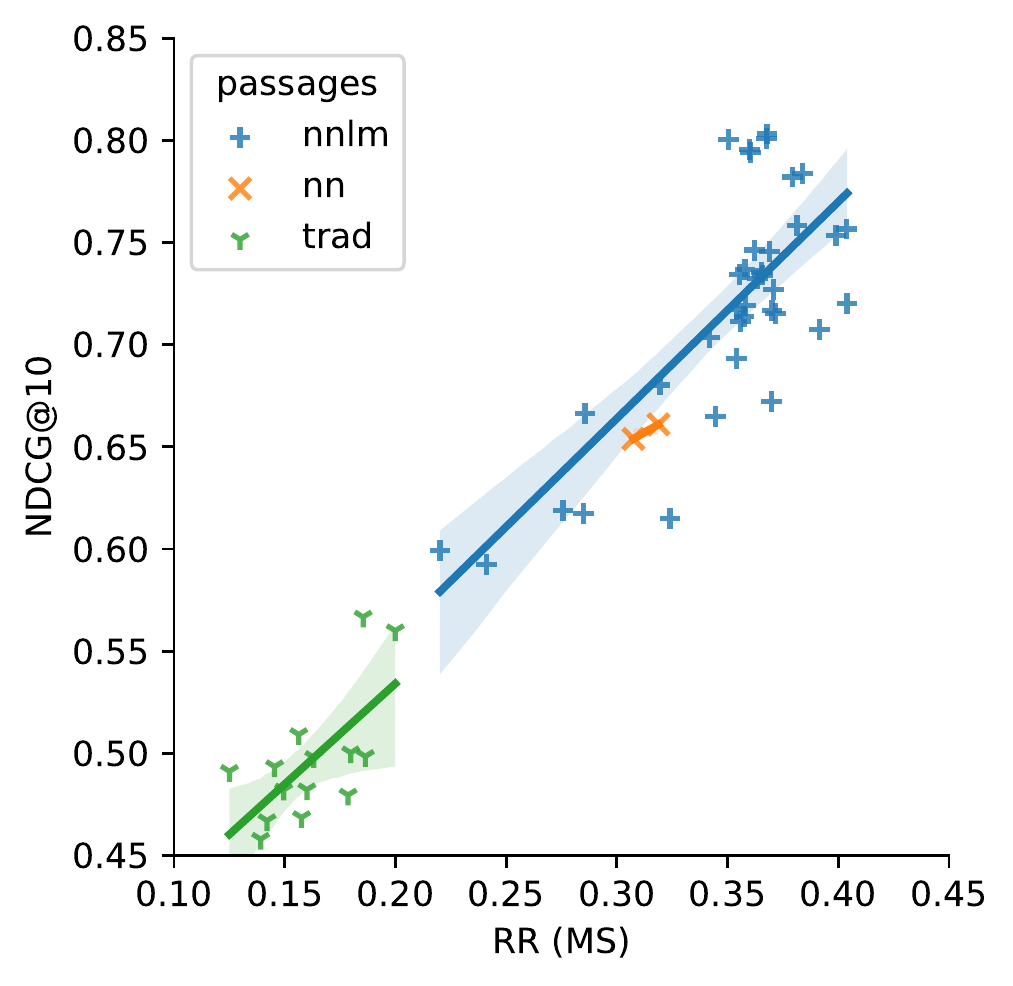}
    \caption{Leaderboard metrics agreement analysis. For document runs, the agreement between the leaderboard metric RR (MS) and the main TREC metric NDCG@10 is lower this year. The Kendall correlation is $\tau=0.46$, compared to $\tau=0.69$ in 2019. For the passage task, we see $\tau=0.69$ in 2020, compared to $\tau=0.68$ in 2019.}
    \label{fig:rrms_vs_ndcg}
\end{figure}

Figure~\ref{fig:rrms_vs_ndcg} shows the agreement between the results using MS MARCO and NIST labels for the document retrieval and passage retrieval tasks. While the agreement between the evaluation setup based on MS MARCO and TREC seems reasonable for both tasks, agreements for the document ranking task seems to be lower (Kendall correlation of $0.46$) than agreements for the passage task (Kendall correlation of $0.69$). This value is also lower than the correlation we observed for the document retrieval task for last year. 

In Table~\ref{tab:kendall_by_tasktype} we show how the agreement between the two evaluation setups varies across task and run type. Agreement on which are the best neural network runs is high, but correlation for document trad runs is close to zero.

\begin{table}[]
    \centering
    \caption{Leaderboard metrics breakdown. The Kendall agreement ($\tau$) of NDCG@10 and RR (MS) varies across task and run type. Agreement on the best neural network runs is high, but agreement on the best document trad runs is very low. We do not list the agreement for passage nn runs since there are only two runs.}
    \begin{tabular}{lrr}
    \toprule
    run type &  docs &  passages \\
    \midrule
    nnlm &  0.83 &      0.76 \\
    nn   &  0.96 &      --- \\
    trad &  0.03 &      0.67 \\
    \midrule
    all  & 0.46 & 0.69 \\
    \bottomrule
    \end{tabular}
    \label{tab:kendall_by_tasktype}
\end{table}

One explanation for this low correlation could be use of the ORCAS dataset. ORCAS was mainly used in the document retrieval task, and could bring search results more in line with Bing's results, since Bing's results are what may be clicked. Since MS MARCO sparse labels were also generated based on top results from Bing, we would expect to see some correlation between ORCAS runs and MS MARCO labels (and Bing results). By contrast, NIST judges had no information about what results were retrieved or clicked in Bing, so may have somewhat less correlation with Bing's results and users.

\begin{figure}
    \centering
    \includegraphics[width=0.49\textwidth]{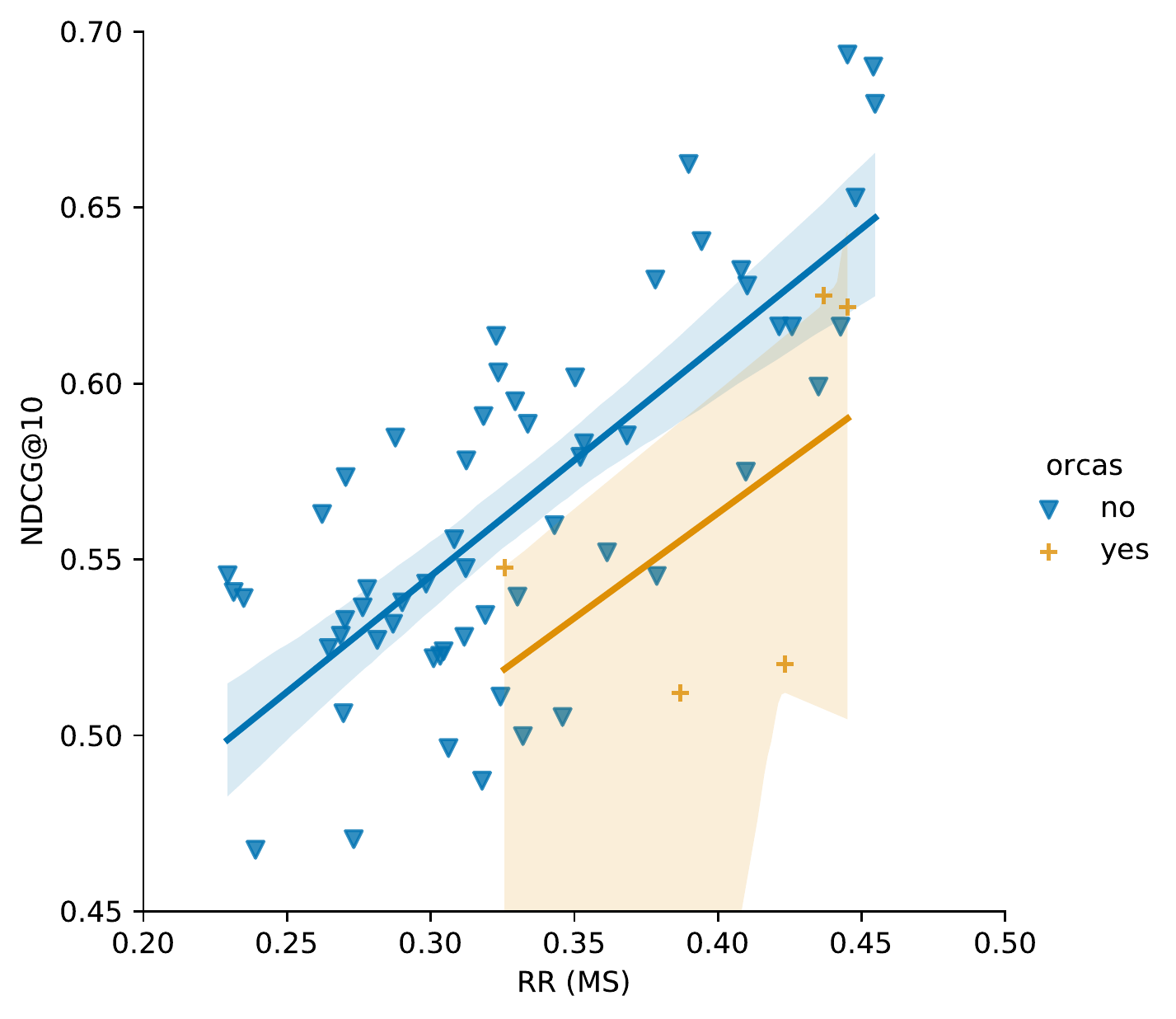}
    \caption{This year it was not necessary to use ORCAS data to achieve the highest NDCG@10. ORCAS runs did somewhat better on the leaderboard metric RR (MS), which uses different labels from the other metrics. This may indicate an alignment between the Bing user clicks in ORCAS with the labeled MS MARCO results, which were also generated by Bing.}
    \label{fig:orcas_scatter}
\end{figure}

In Figure~\ref{fig:orcas_scatter} we compare the results from the two evaluation setups when the runs are split based on the usage of the ORCAS dataset. Our results suggest that runs that use the ORCAS dataset did perform somewhat better based on the MS MARCO evaluation setup. While the similarities between the ORCAS dataset and the MS MARCO labels seem to be one reason for the mismatch between the two evaluation results, it is not enough to fully explain the $0.03$ correlation in Table\ref{tab:kendall_by_tasktype}. Removing the ORCAS ``trad'' runs only increases the correlation to $0.13$. In the future we plan to further analyze the possible reasons for this poor correlation, which could also be related to 1) the different metrics used in the two evaluation setups (RR vs. NDCG@10), 2) the different sensitivity of the datasets due to the different number of queries and number of documents labelled per query), or 3) difference in relevance labels provided by NIST assessors vs. labels derived from clicks. 

\section{Conclusion}
\label{sec:conclusion}
The TREC 2020 Deep Learning Track has provided two large training datasets, for a document retrieval task and a passage retrieval task, generating two ad hoc test collections with good reusability. The main document and passage training datasets in 2020 were the same as those in 2019. In addition, as part of the 2020 track, we have also released a large click dataset, the ORCAS dataset, which was generated using the logs of the Bing search engine.

For both tasks, in the presence of large training data, this year's non-neural network runs were outperformed by neural network runs. While usage of the ORCAS dataset seems to help improve the performance of the systems, it was not necessary to use ORCAS data to achieve the highest NDCG@10. 

We compared reranking approaches to end-to-end retrieval approaches, and in this year's track there was not a huge difference, with some runs performing well in both regimes. This is another result that would be interesting to track in future years, since we would expect that end-to-end retrieval should perform better if it can recall documents that are unavailable in a reranking subtask.

This year the number of runs submitted for both tasks have increased compared to last year. In particular, number of non-neural runs have increased. Hence, test collections generated as part of this year's track may be more reusable compared to last year since these test collections may be fairer towards evaluating the quality of unseen non-neural runs. We note that the number of ``nn'' runs also seems to be smaller this year. We will continue to encourage a variety of approaches in submission, to avoid converging too quickly on one type of run, and to diversify the judging pools.

Similar to last year, in this year's track we have two types of evaluation label for each task. Our official labels are more comprehensive, covering a large number of results per query, and labeled on a four point scale at NIST. We compare this to the MS MARCO labels, which usually only have one positive result per query. While there was a strong correlation between the evaluation results obtained using the two datasets for the passage retrieval task, the correlation for the document retrieval task was lower. Part of this low correlation seems to be related to the usage of the ORCAS dataset (which is generated using similar dataset as the one used to generate the MS MARCO labels) by some runs, and evaluation results based on MS MARCO data favoring these runs. However, our results suggest that while the ORCAS dataset could be one reason for the low correlation, there might be other reasons causing this reduced correlation, which we plan to explore as future work.

\bibliographystyle{plainnat}
\bibliography{bibtex}

\end{document}